\documentclass[12pt]{article}
\usepackage{fullpage}
\usepackage{amsmath}
\usepackage{amssymb}
\usepackage[dvips]{epsfig}

\def\##1{\underline{#1}}
\def\=#1{\underline{\underline{#1}}}

\def\+#1{\underline{\bf #1}}
\def\*#1{\underline{\underline{\bf #1}}}

\def\r#1{(\ref{#1})}
\def\l#1{\label{#1}}
\def\c#1{\cite{#1}}

\def\le{\left(}
\def\ri{\right)}
\def\les{\left[}
\def\ris{\right]}
\def\lec{\left\{}
\def\ric{\right\}}

\def\.{\mbox{ \tiny{$^\bullet$} }}

\def\epso{\epsilon_{\scriptscriptstyle 0}}

\def\muo{\mu_{\scriptscriptstyle 0}}

\def\eps{\epsilon}

\begin{document}

\begin{center}

{\bf {\Large Anisotropic enhancement of group velocity in a
homogenized dielectric composite medium }}

 \vspace{10mm} \large

Tom G. Mackay\footnote{Corresponding author. E--mail: T.Mackay@ed.ac.uk}\\
{\em School of Mathematics,
University of Edinburgh, Edinburgh EH9 3JZ, UK}\\
\bigskip
 Akhlesh  Lakhtakia\footnote{E--mail: akhlesh@psu.edu; also
 affiliated with Department of Physics, Imperial College, London SW7 2 AZ, UK}\\
 {\em CATMAS~---~Computational \& Theoretical
Materials Sciences Group\\ Department of Engineering Science and
Mechanics\\ Pennsylvania State University, University Park, PA
16802--6812, USA}\\

\end{center}
\vspace{4mm}

\normalsize

\begin{abstract}
Under certain circumstances, the group velocity in a homogenized
composite medium (HCM) can exceed the group velocity in its
component material phases. We explore this phenomenon for a
uniaxial dielectric HCM comprising isotropic component material
phases distributed as oriented spheroidal particles. The theoretical
approach is based upon the Bruggeman homogenization formalism.
Enhancement in group velocity in the HCM
with respect to the component material phases is shown to be sensitively
dependent upon the shape of the component spheroids and their
alignment relative to the direction of propagation.

\end{abstract}



\section{Introduction}

The process of homogenization involves the combination of two (or
more) component material phases to produce a single, effectively
homogeneous, composite medium \c{Ward,Neel,LOCM}. Typically, the
constitutive properties
of the component material phases are relatively simple as compared with
those of the homogenized composite medium (HCM).
Through homogenization, novel and potentially useful material
properties may be realized \c{Walser, M03}. Many examples of  material properties
being extended~---~or indeed  entirely new material properties
being realized~---~as a result of homogenization can be found
within the regimes of linear and nonlinear electromagnetics
\c{M05}.

 An interesting   result concerns the
electromagnetic group velocity in HCMs. Under certain
circumstances,
 the group velocity in an HCM can exceed the group velocities in its
 component material phases. This issue has been investigated for
 isotropic dielectric composite mediums using the Maxwell Garnett \c{Solna1,Solna2}
  and the Bruggeman  \c{ML2004} homogenization formalisms.
In these studies,  an enhancement in
 group velocity  is demonstrated through homogenizing two
 component material phases, one of which
is characterized by a relatively large permittivity and relatively
small frequency--dispersive term as compared with the other
component material phase.

Enhancement of
group velocity in a laminate composite medium has been
considered by using a volume--weighted sum to
 estimate the HCM permittivity \c{Solna1,Solna2}.
The directional properties of group--velocity enhancement are
further explored in this communication. Specifically, we consider
a uniaxial dielectric HCM which develops from the homogenization
of a random assembly of oriented spheroidal particles. The
component material phases  are themselves electromagnetically
isotropic. Our theoretical analysis is founded upon the Bruggeman
homogenization formalism \c{Michel}.

\section{Homogenization}

Let us consider the homogenization of a composite medium containing two
component material  phases, labelled as $a$ and $b$. Both
component material phases are taken to be isotropic dielectric
mediums: $\eps^a$ and $\eps^b$ denote the permittivity scalars of
phases $a$ and $b$, respectively. In order to focus in particular
upon the  phenomenon of  enhancement of group velocity, without
being distracted by the complications arising from dielectric
loss, the component material phases are assumed to be
nondissipative; i.e., $\eps^{a,b} \in \mathbb{R}$. The component
material phases are envisioned as random distributions of
identically oriented, spheroidal particles. The spheroidal
shape~---~which is taken to be the same for all particles of
phases $a$ and $b$~---~is parameterized via the shape dyadic
\begin{equation}
 \=U =  U_\perp \=I + \le  U_\parallel - U_\perp \ri \, \hat{\#c} \, \hat{\#c}\,,
\end{equation}
where $\=I$ is the identity 3$\times$3 dyadic and the unit vector
$\hat{\#c}$ is parallel to the spheroid's axis of rotational
symmetry. The spheroid's surface is described by the vector
\begin{equation}
\#r_{\,s} (\theta, \phi) = \eta \, \=U \. \hat{\#r} (\theta,
\phi)\,,
\end{equation}
with $ \hat{\#r} $ being the radial unit vector from the spheroid's
centroid and specified by the
spherical polar coordinates $\theta$ and $\phi$. The linear
dimensions of the spheroid, as determined by the parameter $\eta$,
are  assumed to be small relative to the electromagnetic
wavelength(s).

The permittivity dyadic of the resulting HCM,
\begin{equation}
\=\eps^{Br} =  \eps^{Br}_\perp \=I + \le \eps^{Br}_\parallel -
\eps^{Br}_\perp \ri\, \hat{\#c} \, \hat{\#c}, \l{eps_Br}
\end{equation}
is estimated using the Bruggeman homogenization formalism as the
solution of the equation
\begin{equation}
f_a \, \=a^{a} + f_b \, \=a^{b}  = \=0\,, \l{Br}
\end{equation}
where $f_a$ and $f_b = 1 - f_a$ denote the respective volume
fractions of the material component phases $a$ and $b$ \c{Michel}.
The polarizability dyadics in \r{Br} are defined as
\begin{equation}
\=a^{\ell } = \le \eps^\ell \=I - \=\eps^{Br} \ri \.\les \, \=I +
 \=D \. \le \eps^\ell \=I - \=\eps^{Br} \ri \,\ris^{-1}, \qquad
(\ell = a,b), \l{polar}
\end{equation}
wherein the depolarization dyadic is given by the surface integral
 \c{LW94,SL96,LMW97}
\begin{equation}
\=D = \frac{1}{ 4 \pi} \, \int^{2 \pi}_0 \, d \phi \, \int^\pi_0
\, d \theta \, \sin \theta \, \le \frac{1}{
\hat{\#r}\.\=U^{-1}\.\=\eps^{Br}\.\=U^{-1}\.\hat{\#r}} \ri
\=U^{-1}\.\hat{\#r} \, \hat{\#r} \. \=U^{-1}\,. \l{depol}
\end{equation}

The  depolarization dyadic may
be expressed as
\begin{equation}
\=D = D_\perp\, \=I + \le D_\parallel - D_\perp \ri \hat{\#c}
\,\hat{\#c}\,,
\end{equation}
where
\begin{eqnarray}
D_\parallel &=& \frac{\gamma}{ \eps^{Br}_\parallel } \, \Gamma_\parallel ( \gamma ), \l{Dx}\\
D_\perp&=& \frac{1}{ \eps^{Br}_\perp} \, \Gamma_\perp (\gamma),
\l{D}
\end{eqnarray}
The terms $\Gamma_\parallel$ and $\Gamma_\perp$ herein are
functions of the real--valued parameter
\begin{equation}
\gamma = \frac{U^2_\perp \eps^{Br}_\parallel}{U^2_\parallel
\eps^{Br}_\perp}\,;
\end{equation}
they have the  representations
\begin{eqnarray}
\Gamma_\parallel (\gamma) &=& \frac{1}{4 \pi}\, \int^{2 \pi}_0 \,
d \phi \, \int^\pi_0 \, d \theta \, \frac{\cos^2 \phi \sin^3
\theta}{\cos^2 \theta + \sin^2 \theta \le \gamma \cos^2 \phi +
\sin^2 \phi \ri}, \l{dx}\\
\Gamma_\perp(\gamma) &=&  \frac{1}{4 \pi}\, \int^{2 \pi}_0 \, d
\phi \, \int^\pi_0 \, d \theta \, \frac{\sin^2 \phi \sin^3
\theta}{\cos^2 \theta + \sin^2 \theta \le \gamma \cos^2 \phi +
\sin^2 \phi \ri}. \l{d}
\end{eqnarray}
The surface integrals   \r{dx} and \r{d} may be evaluated as
\begin{eqnarray}
\Gamma_\parallel (\gamma )&=& \left\{
\begin{array}{lcr}
\displaystyle{
  \frac{\sinh^{-1} \sqrt{\frac{1
-\gamma}{\gamma} }}{\le 1 - \gamma  \ri^{\frac{3}{2}}} -
\frac{1}{1-\gamma }} && \hspace{14mm} \mbox{for} \;\; 0 < \gamma <
1
\\ & & \\
\displaystyle{ \frac{1}{\gamma - 1} - \frac{\sec^{-1}
\sqrt{\gamma} } {\le \gamma - 1 \ri^{\frac{3}{2}}}}& & \mbox{for}
\;\; \gamma > 1
\end{array}
\right., \\
\Gamma_\perp( \gamma )&=& \left\{
\begin{array}{lcr}
\displaystyle{ \frac{1}{2} \le  \frac{1}{1-\gamma }-
  \frac{ \gamma \sinh^{-1} \sqrt{\frac{1
-\gamma}{\gamma} }}{\le 1 - \gamma  \ri^{\frac{3}{2}}} \ri } &&
\mbox{for} \;\; 0 < \gamma < 1
\\ & & \\
\displaystyle{\frac{1}{2} \le \frac{\gamma \sec^{-1} \sqrt{\gamma}
} {\le \gamma - 1 \ri^{\frac{3}{2}}} - \frac{1}{\gamma - 1}  \ri
}& & \mbox{for} \;\; \gamma > 1
\end{array}
\right..
\end{eqnarray}
We exclude the cases of
\begin{itemize}
\item the isotropic HCM with $\gamma = 1 $, and
\item
the anomalous hyperbolic HCM with $\gamma < 0$ \c{MLD2005}
\end{itemize}
 from
consideration.

The dyadic Bruggeman equation \r{Br} provides the two nonlinear
scalar equations
\begin{eqnarray}
&& \frac{\eps^a - \eps^{Br}_\parallel }{1 + D_\parallel \le \eps^a
- \eps^{Br}_\parallel \ri} f_a + \frac{\eps^b -
\eps^{Br}_\parallel }{1 + D_\parallel \le \eps^b -
\eps^{Br}_\parallel \ri} f_b = 0 \,, \l{Br_1}
\\
&& \frac{\eps^a - \eps^{Br}_\perp}{1 + D_\perp\le \eps^a -
\eps^{Br}_\perp\ri} f_a + \frac{\eps^b - \eps^{Br}_\perp}{1 +
D_\perp\le \eps^b - \eps^{Br}_\perp\ri} f_b = 0\,, \l{Br_2}
\end{eqnarray}
coupled via $D_{\perp, \parallel}$,  which can be solved
straightforwardly for $\eps^{Br}_\parallel$ and $\eps^{Br}_\perp$
using standard numerical techniques.

\section{Group velocity}

Let us consider a wavepacket which is a superposition of
planewaves with phasors
\begin{equation}
\left.\begin{array}{l}
\#E(\#r) = \#E_0\, \exp \le i \#k \. \#r  \ri  \\[5pt]
\#H(\#r) = \#H_0\, \exp \le i \#k \. \#r \ri
\end{array}\right\}.
\l{pw}
\end{equation}
The group velocity  $\#v_g$ of the wavepacket is conventionally defined in terms of the
gradient of the angular frequency $\omega$
with respect to $\#k$
 \c{Chen}; i.e.,
\begin{equation}
\#v_g = \left. \nabla_{\#k}\, \omega \, \right|_{\omega = \omega
(k_{avg})}  \l{vg_Def}\,,
\end{equation}
where $k_{avg}$ denotes the average wavenumber of the wavepacket.
Herein we adopt the compact notation
\begin{equation}
\nabla_{\#k} \equiv \le \frac{\partial }{\partial k_x } , \,
\frac{\partial }{\partial k_y } , \, \frac{\partial }{\partial k_z
} \, \ri \end{equation} for the gradient operator with respect to
$\#k$, where  $\le k_x, k_y, k_z \ri$ is the representation of
$\#k$ in terms of its Cartesian components.

In order to calculate the group velocity in the uniaxial
dielectric HCM  \r{eps_Br}, denoted as $\#v^{Br}_g$,  we exploit
the corresponding planewave dispersion relation as follows. The
combination of \r{eps_Br} with the source--free Maxwell curl
postulates
\begin{equation}
\left.\begin{array}{l}
\nabla \times \#E(\#r) = i \omega \#B ( \#r )\\[5pt]
\nabla \times \#H(\#r) = - i \omega \#D ( \#r )
\end{array}\right\},
\l{Maxwell}
\end{equation}
delivers
 the vector Helmholtz equation
\begin{equation}
\les \,\le \nabla \times \=I \,\ri \. \le \nabla \times \=I\, \ri
- \muo \omega^2 \=\eps^{Br} \,\ris \.  \#E_0  = \#0\,,
\l{Helmholtz}
\end{equation}
with $\muo$ being the permeability of free space. The requirement
that  \r{Helmholtz} provide nonzero solutions for the planewave
phasors \r{pw} yields the  dispersion relation
\begin{equation}
W( \#k, \omega ) = 0\,, \l{disp}
\end{equation}
wherein the scalar function $W$ is defined as
\begin{equation}
W( \#k, \omega ) =  \le \#k \. \#k  - \eps^{Br}_\perp\muo \omega^2
\ri \le \#k \. \=\eps^{Br} \. \#k  - \eps^{Br}_\parallel
\eps^{Br}_\perp\muo \omega^2 \ri \,. \l{W}
\end{equation}
The dispersion relation \r{disp} admits two wavevector solutions:
the ordinary wavevector $\#k_{or}$ and the extraordinary wavector
$\#k_{ex}$, satisfying
\begin{equation}
\left.
\begin{array}{l}
\#k_{or} \. \#k_{or}  - \eps^{Br}_\perp\muo \omega^2 =0 \\[5pt]
\#k_{ex} \. \=\eps^{Br} \. \#k_{ex}  - \eps^{Br}_\parallel
\eps^{Br}_\perp\muo \omega^2 =0
\end{array}
\right\}. \l{k_or_ex}
\end{equation}
We note that the magnitude of the
ordinary wavevector is direction--independent, and the
ordinary and extraordinary wavevectors coincide when
$\#k_{ex}$ is directed along $\hat{\#c}$.

 By taking the gradient of the dispersion relation
\r{disp} with respect to $\#k$,  we find
\begin{equation}
\nabla_{\#k} W + \frac{\partial W}{\partial \omega} \,
\nabla_{\#k} \omega = \#0\,.
\end{equation}
Hence,  the HCM group velocity  \r{vg_Def} may be expressed as
\begin{equation}
\#v^{Br}_g = - \left. \frac{ 1}{\partial W /
\partial \omega }
\nabla_{\# k}  W \,
 \right|_{\omega = \omega (k_{avg})} \l{vg}\,.
\end{equation}
The partial derivative terms involving $W$ are found to be
\begin{eqnarray}
\nabla_{\# k}  W &=& 2 \les \le \#k \. \=\eps^{Br} \. \#k -
\omega^2 \muo \eps^{Br}_\perp\eps^{Br}_\parallel \ri \#k + \le
\#k\.\#k - \omega^2 \muo \eps^{Br}_\perp\ri \=\eps^{Br} \. \#k
\,\ris , \l{dWdk}
\\
\frac{\partial W}{\partial \omega } &=& \le \#k \. \#k - \omega^2
\muo \eps^{Br}_\perp\ri \lec  \#k \. \frac{d \=\eps^{Br}}{d
\omega} \. \#k - \muo \omega  \les 2
\eps^{Br}_\perp\eps^{Br}_\parallel + \omega \le \frac{d
\eps^{Br}_\perp}{d \omega} \eps^{Br}_\parallel + \eps^{Br}_\perp\frac{d \eps^{Br}_\parallel}{d \omega} \ri \ris \ric \nonumber \\
&& - \muo \omega \le 2  \eps^{Br}_\perp+ \omega \frac{d
\eps^{Br}_\perp}{d \omega} \ri \le \#k \. \=\eps^{Br} \. \#k  -
\eps^{Br}_\parallel \eps^{Br}_\perp \muo \omega^2 \ri \,, \l{dWdw}
\end{eqnarray}
with
\begin{equation}
\frac{d \=\eps^{Br}}{d \omega}  =
 \frac{d \eps^{Br}_\perp}{d \omega}\, \=I + \le  \frac{d \eps^{Br}_\parallel}{d \omega} -
 \frac{d \eps^{Br}_\perp}{d \omega} \ri\, \hat{\#c} \, \hat{\#c}
\,. \l{deps_Br}
\end{equation}
By virtue of \r{k_or_ex}, we see that the ordinary and the
extraordinary group velocities are
given by
\begin{equation}
\left. \#v^{Br}_g \right|_{\#k = \#k_{or}} =
\frac{2}{\displaystyle{\omega \muo \le 2 \eps^{Br}_\perp+ \omega
\frac{d \eps^{Br}_\perp}{d \omega} \ri}} \; \#k_{or} \l{vg_or}
\end{equation}
and
\begin{equation}
\left. \#v^{Br}_g \right|_{\#k = \#k_{ex}} = \frac{2}{
\displaystyle{ \omega \muo \les
 2 \eps^{Br}_\perp\eps^{Br}_\parallel + \omega \le \frac{d \eps^{Br}_\perp}{d
\omega} \eps^{Br}_\parallel + \eps^{Br}_\perp\frac{d
\eps^{Br}_\parallel}{d \omega} \ri \ris -  \#k_{ex} \. \frac{d
\=\eps^{Br}}{d \omega} \. \#k_{ex} }} \; \=\eps^{Br} \.
\#k_{ex}\,, \l{vg_ex}
\end{equation}
respectively.

In order to find the derivatives of $\eps^{Br}_\perp$ and
$\eps^{Br}_\parallel$ needed to evaluate the group velocities
\r{vg_or} and \r{vg_ex}, we have to exploit the Bruggeman equations
\r{Br_1} and \r{Br_2}. As a precursor, let us first note the
derivatives of the depolarization dyadic components
\begin{eqnarray}
\frac{d D_\parallel}{d \omega } &=&  \alpha_{11} \frac{d
\eps^{Br}_\parallel}{d
\omega} + \alpha_{12} \frac{d \eps^{Br}_\perp}{d \omega}\,,\\
\frac{d D_\perp}{d \omega}  &=& \alpha_{21} \frac{d
\eps^{Br}_\parallel}{d \omega} + \alpha_{22} \frac{d
\eps^{Br}_\perp}{d \omega}\,,
\end{eqnarray}
with
\begin{eqnarray}
\alpha_{11} &=& \frac{U^2_\perp}{U^2_\parallel \eps^{Br}_\parallel
\eps^{Br}_\perp} \le \Gamma_\parallel + \gamma
\frac{d\Gamma_\parallel}{d \gamma} \ri - \frac{\gamma
\Gamma_\parallel}{\le
\eps^{Br}_\parallel \ri^2} \,,\\
\alpha_{12} &=& - \frac{U^2_\perp }{U^2_\parallel \le \eps^{Br}_\perp\ri^2} \le \Gamma_\parallel + \gamma \frac{d\Gamma_\parallel}{d \gamma} \ri\,,\\
\alpha_{21} &=& \le \frac{U^2_\perp}{U^2_\parallel \le
\eps^{Br}_\perp\ri^2} \ri \,
\frac{d\Gamma_\perp}{d \gamma}  \,,\\
\alpha_{22} &=& - \le \frac{U^2_\perp
\eps^{Br}_\parallel}{U^2_\parallel \le \eps^{Br}_\perp\ri^3} \ri\,
\frac{d\Gamma_\perp}{d \gamma}  - \frac{ \Gamma_\perp}{\le
\eps^{Br}_\perp\ri^2} \,,
\end{eqnarray}
and
\begin{eqnarray}
\frac{d\Gamma_\parallel}{d \gamma} &=&
 \left\{
\begin{array}{lcr}
\displaystyle{ \frac{1}{2} \le
  \frac{3 \sinh^{-1} \sqrt{\frac{1
-\gamma}{\gamma} }}{\le 1 - \gamma  \ri^{\frac{5}{2}}} -
\frac{1 + 2 \gamma}{ \le 1-\gamma \ri^2 \gamma } \ri } && \hspace{10mm} \mbox{for} \;\; 0 < \gamma <
1
\\ & & \\
\displaystyle{\frac{1}{2} \le  - \frac{1 + 2 \gamma}{\le \gamma - 1 \ri^2 \gamma } + \frac{3 \sec^{-1}
\sqrt{\gamma} } {\le \gamma - 1 \ri^{\frac{5}{2}}} \ri }& & \mbox{for}
\;\; \gamma > 1
\end{array}
\right., \\
\frac{d \Gamma_\perp}{d \gamma} &=& \left\{
\begin{array}{lcr}
\displaystyle{ \frac{1}{4} \le   \frac{3}{ \le 1-\gamma \ri^2 } -
  \frac{\le 2 +  \gamma \ri \sinh^{-1} \sqrt{\frac{1
-\gamma}{\gamma} }}{\le 1 - \gamma  \ri^{\frac{5}{2}}} \ri } &&
\mbox{for} \;\; 0 < \gamma < 1
\\ & & \\
\displaystyle{\frac{1}{4} \le - \frac{ \le 2+ \gamma \ri \sec^{-1} \sqrt{\gamma}
} {\le \gamma - 1 \ri^{\frac{5}{2}}} + \frac{3}{\le \gamma - 1 \ri^2}  \ri
}& & \mbox{for} \;\; \gamma > 1
\end{array}
\right..
\end{eqnarray}
Now we turn to the Bruggeman equations \r{Br_1} and \r{Br_2}. Their
derivatives with respect to $\omega$  may be written  as
\begin{eqnarray}
&&  \beta_{11} \frac{d \eps^{Br}_\parallel}{d
\omega} + \beta_{12} \frac{d \eps^{Br}_\perp}{d \omega} + \beta_{13} = 0\,,\\
&& \beta_{21} \frac{d \eps^{Br}_\parallel}{d \omega} + \beta_{22}
\frac{d \eps^{Br}_\perp}{d \omega} + \beta_{23} = 0 \,,
\end{eqnarray}
with
\begin{eqnarray}
\beta_{11} &=& \alpha_{11} \le \eps^a - \eps^{Br}_\parallel \ri
\le \eps^b - \eps^{Br}_\parallel \ri + D_\parallel \le 2
\eps^{Br}_\parallel - \eps^a - \eps^b \ri -1 \,, \\
\beta_{12} &=&
\alpha_{12} \le \eps^a - \eps^{Br}_\parallel \ri \le \eps^{b} - \eps^{Br}_\parallel \ri \,,\\
\beta_{13} &=& \les f_a + D_\parallel \le \eps^b -
\eps^{Br}_\parallel \ri \ris\, \frac{ d \eps^a}{d \omega} + \les
f_b + D_\parallel \le \eps^a - \eps^{Br}_\parallel \ri \ris\,
\frac{ d \eps^b}{d \omega}\,, \\
\beta_{21} &=&
\alpha_{21} \le \eps^a - \eps^{Br}_\perp\ri \le \eps^{b} - \eps^{Br}_\perp\ri \,,\\
\beta_{22} &=& \alpha_{22} \le \eps^a - \eps^{Br}_\perp \ri \le
\eps^b - \eps^{Br}_\perp \ri + D_\perp \le 2
\eps^{Br}_\perp - \eps^a - \eps^b \ri -1 \,, \\
\beta_{23} &=& \les f_a + D_\perp \le \eps^b - \eps^{Br}_\perp \ri
\ris\, \frac{ d \eps^a}{d \omega} + \les f_b + D_\perp \le \eps^a
- \eps^{Br}_\perp \ri \ris\, \frac{ d \eps^b}{d \omega}
 \,.
\end{eqnarray}
The  derivatives of  $\eps^{Br}_\perp$ and $\eps^{Br}_\parallel$
therefore finally emerge as
\begin{eqnarray}
 \frac{d \eps^{Br}_\parallel}{d
\omega} &=&  \frac{ \beta_{12} \beta_{23} - \beta_{22}
\beta_{13}}{\beta_{11} \beta_{22} - \beta_{12} \beta_{21}}
\,, \l{dexdw} \\
 \frac{d \eps^{Br}_\perp}{d
\omega} &=&  \frac{ \beta_{21} \beta_{13} - \beta_{11}
\beta_{23}}{\beta_{11} \beta_{22} - \beta_{12} \beta_{21}} \,.
\l{dedw}
\end{eqnarray}

To summarize,
given a  uniaxial dielectric HCM with permittivity dyadic $\=\eps^{Br}$
estimated using the Bruggeman homogenization formalism,
 the group velocity \r{vg_Def}  may be computed using the expression \r{vg}, with
\r{dWdk} and \r{dWdw}, wherein the derivatives of
$\eps^{Br}_\perp$ and $\eps^{Br}_\parallel$ are provided by
\r{dexdw} and \r{dedw}.

\section{Numerical studies}
Without loss of generality, let us choose the  axis of rotational
symmetry of the component spheroids to lie along the $x$ axis,
i.e., $\hat{\#c} = \hat{\#x}$. We consider wavevectors lying in
the $xy$ plane, oriented at an angle $\theta$ to the $x$ axis.
That is, we take
\begin{equation} \#k = k \lec \cos \theta \,, \sin \theta \,, 0 \ric\,.
\end{equation}
Thus, the magnitudes $k = k_{or} \equiv | \#k_{or} |$ and $k =
k_{ex} \equiv | \#k_{ex} |$ of the ordinary and extraordinary
wavevectors arise from \r{k_or_ex} as \c{Chen}
\begin{eqnarray}
 k_{or} &=& \omega \sqrt{\muo \eps^{Br}_\perp} \,,\\
k_{ex}& =& \omega \sqrt{ \frac{\muo \eps^{Br}_\parallel
\eps^{Br}_\perp }{\eps^{Br}_\parallel \cos^2 \theta +
\eps^{Br}_\perp\sin^2 \theta}}\,.
\end{eqnarray}

Let us explore numerically  the enhancement in group velocity that
can arise through homogenization, paying special attention to
directional effects induced by the shape of the component
spheroidal particles. In particular, we  choose the component
material phase $a$ to have a relatively high permittivity $\eps^a$
and a relatively small frequency--dispersion term $d \eps^a / d \omega$,
compared with the component material phase $b$. As representative
constitutive parameter values, we set: $\eps^a = 30 \epso$, $
\left. \le d \eps^a / d \omega \ri \right|_{\omega = \omega_o} =
6\epso / \omega_o$, $\eps^b =  1.2 \epso$ and  $ \left. \le d
\eps^b / d \omega \ri \right|_{\omega = \omega_o} = 12\epso /
\omega_o$, where $\epso$ is the permittivity of free space.

In Figure~1, the Bruggeman estimates of the HCM permittivity
parameters $\eps^{Br}_\parallel$ and $\eps^{Br}_\perp$ are plotted
as functions of volume fraction $f_a$, for the range of values of
$\rho = U_\parallel / U_\perp$ shown in Table~1. Clearly,
$\eps^{Br}_{\perp, \parallel} \rightarrow \eps^b$ as $f_a
\rightarrow 0$ and $\eps^{Br}_{\perp, \parallel} \rightarrow
\eps^a$ as $f_a \rightarrow 1$. We see that $\eps^{Br}_\parallel$
becomes an increasingly nonlinear function of $f_a$ as $\rho$
decreases, whereas $\eps^{Br}_\perp$ becomes an increasingly
nonlinear function of $f_a$ as $\rho$ increases.

In Figure~2, the magnitude of the group velocity $v^{Br}_g = |
\#v^{Br}_g |$ of a wavepacket in the chosen HCM is plotted against
volume fraction. The group velocities are calculated with $\#k =
\#k_{ex}$ for $\theta = 0^\circ, 30^\circ, 60^\circ$ and
$90^\circ$. The corresponding
graphs for $ 180^\circ - \theta$ are the same as
those for $\theta$. Since the ordinary wavevector  $\#k_{or} =
\#k_{ex}$ at $\theta = 0^\circ$, the ordinary group velocities for any
$\theta$ are  identical to those provided in
Figure~2(a) wherein the results for $\theta = 0^\circ$ are
presented. The group velocity magnitudes for the component
material phases $a$ and $b$ are $v^a_g = 0.166 c$ and $v^b_g =
0.152 c$, respectively (as is confirmed in Figure~2 by the group
velocity values at $f_a=1$ and $f_a = 0$, respectively), where $c
= 1/\sqrt{\epso \muo}$. Hence, for this particular homogenization
example, group--velocity enhancement arises when $v^{Br}_g >
\mbox{max} \lec v^a_g, v^b_g \ric = 0.166 c$. The
group--velocity--enhancement region is identified by shading in
Figure~2.

 It may be discerned from
Figure~2(a) that group--velocity enhancement occurs  over an
increasingly large range of $f_a$ values as $\rho$ decreases.
Furthermore, the degree of enhancement at $\rho = 20$ is much
smaller than it is at $\rho = 0.05$.

As $\theta$ increases, the range of $f_a$ values at which
group--velocity enhancement occurs progressively decreases for small
values of $\rho$. In fact, at $\theta = 60^\circ$ there is no
longer any enhancement in group velocity for $\rho = 0.05$. At
$\theta = 90^\circ$, the group--velocity enhancement
characteristics at low and high values of $\rho$ are approximately
the reverse of  their respective characteristics at $\theta =
0^\circ$. That is, group--velocity enhancement occurs over a wide
range of $f_a$ values for high values of $\rho$ at $\theta =
90^\circ$, but there is no enhancement in group velocity at low
values of $\rho$.

Clearly therefore, enhancement of group velocity is maximum in a
direction parallel to the longest semi--axis of the spheroidal
particles, which can be prolate ($\rho < 1$) or oblate ($\rho
> 1)$. For spherical particles ($\rho=1$), group--velocity
enhancement is direction--independent, and we recover the results
of the predecessor study \c{ML2004}.

\section{Concluding remarks}

The enhancement in group velocity brought about by homogenization
is  sensitively dependent upon directional  properties. Both the
shape of the component   spheroidal particles, and their
orientation relative to the direction of propagation,  strongly
influence the group--velocity enhancement.

The homogenization scenario presented here deals with the
conceptualization of a uniaxial HCM as arising from identically oriented
 spheroidal particles of
isotropic
component material phases. The homogenization of two uniaxial dielectric
component phases distributed as  spherical
particles is
mathematically equivalent, provided that the distinguished axes of
the component material phases have the same orientation \c{MW2001}.

\vspace{5mm}

\noindent {\bf Acknowledgement:} We thank Prof. G.W. Milton for
drawing our attention to  group--velocity enhancement in isotropic
dielectric composite mediums with HCM permittivity estimated using the
Maxwell Garnett formalism, as described in Refs.
\cite{Solna1} and \cite{Solna2}.

\newpage

\begin{figure}[ht!] \centering \psfull  \epsfig{file=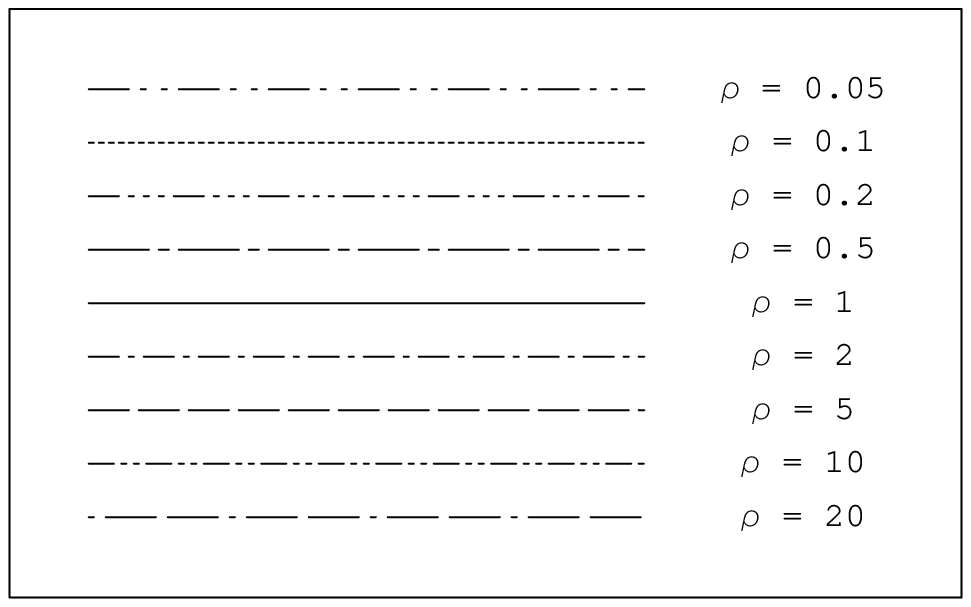,width=5.2in}
\end{figure}
\begin{center}
 Table~1. Key for the values of $\rho = U_\parallel / U_\perp $ used in Figures~1 and 2.
\end{center}

\newpage

\setcounter{figure}{0}

\begin{figure}[!ht]
\centering \psfull \epsfig{file=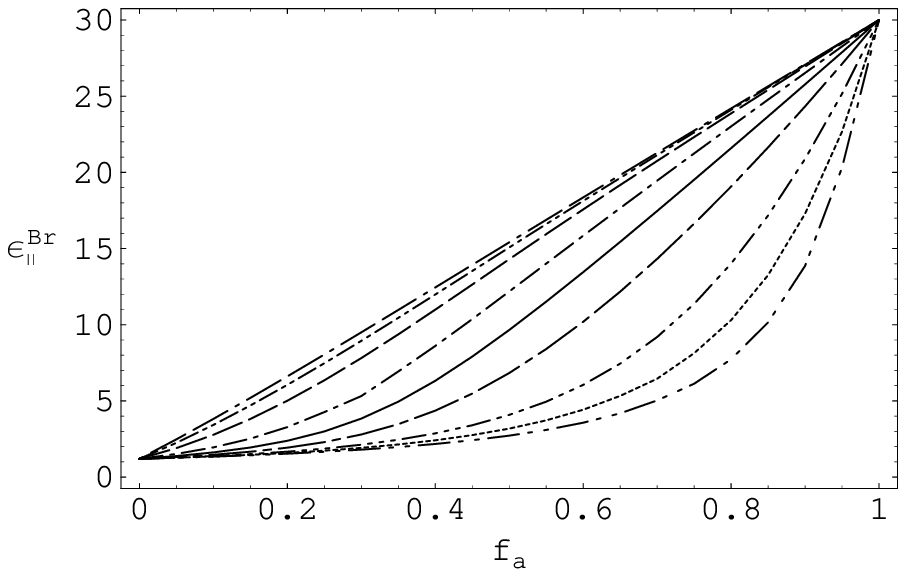,width=5.0in}\\
\epsfig{file=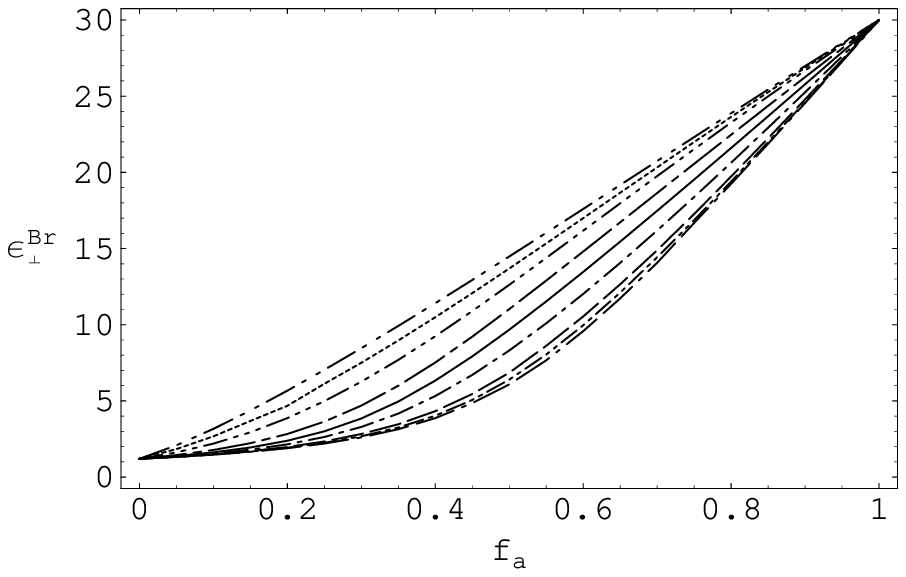,width=5.0in}
  \caption{\label{fig1}
The HCM relative permittivity parameters $\eps^{Br}_\parallel
/\epso$  (above) and $\eps^{Br}_\perp / \epso$ (below) plotted
against volume fraction $f_a$. Permittivities of component
material phases: $\eps_a = 30 \epso$ and $\eps_b = 1.2 \epso $. A
key for the $\rho=U_\parallel / U_\perp$ values is given in  Table~1.
 }
\end{figure}

\newpage

\begin{figure}[!ht]
\centering \psfull \epsfig{file=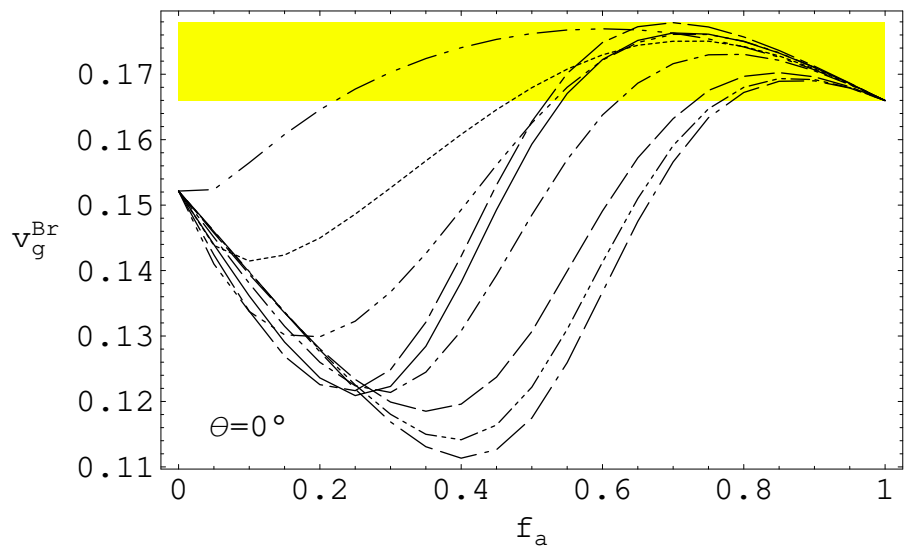,width=5.0in}
  \caption{\label{fig2a}
The magnitude of the HCM group velocity $v^{Br}_g = | \#v^{Br}_g |
$, as estimated using the Bruggeman formalism, plotted against
volume fraction $f_a$. The group velocity is normalized with
respect to $c=1/\sqrt{\epso\muo}$. Constitutive parameters of
component material phases: $\eps^a = 30 \epso$, $ \left. \le d
\eps^a / d \omega \ri \right|_{\omega = \omega_o} = 6 \epso /
\omega_o$, $\eps^b = 1.2 \epso$ and  $ \left. \le d \eps^b / d
\omega \ri \right|_{\omega = \omega_o} = 12  \epso/ \omega_o$. A
key for the $\rho=U_\parallel / U_\perp$ values is given in
Table~1. Shading indicates the region of group--velocity
enhancement. (a) Extraordinary wavevector angle $\theta =
0^\circ$.}\end{figure}

\setcounter{figure}{1}

\begin{figure}[!ht]
\centering \psfull  \epsfig{file=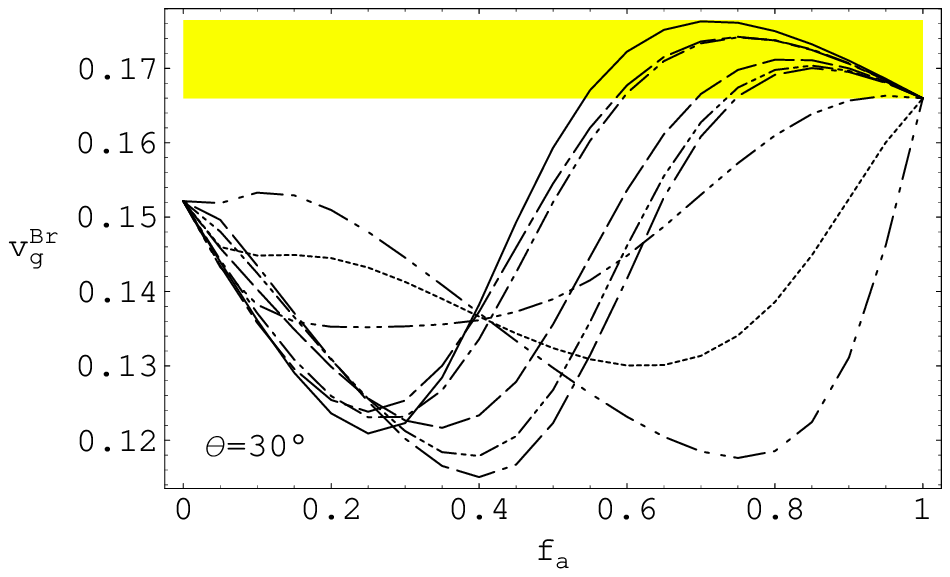,width=5.0in}
  \caption{\label{fig2b}
(b) Extraordinary wavevector angle $\theta = 30^\circ$.}
\end{figure}

\newpage

\setcounter{figure}{1}

\begin{figure}[!ht]
\centering \psfull  \epsfig{file=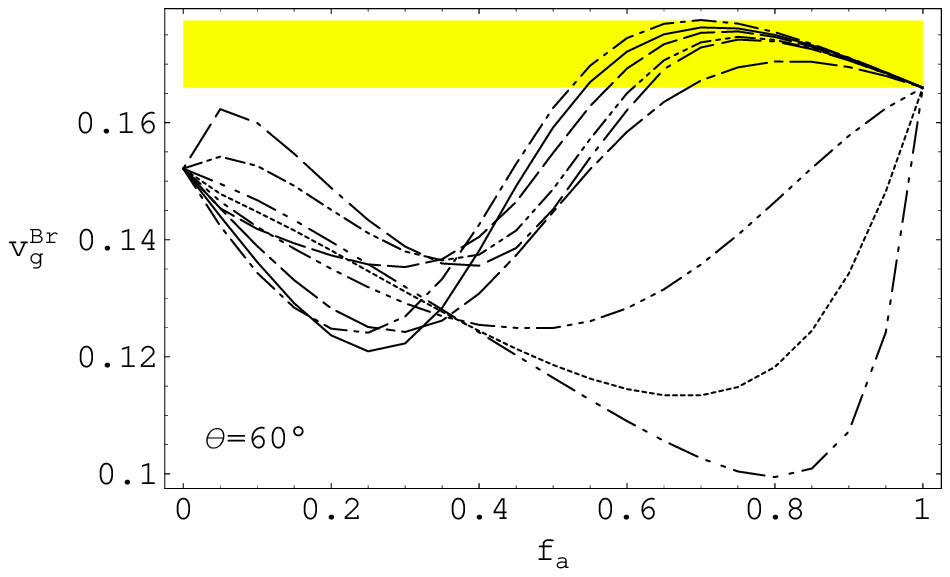,width=5.0in}
  \caption{\label{fig2c}
(c) Extraordinary wavevector angle $\theta = 60^\circ$.}
\end{figure}

\setcounter{figure}{1}

\begin{figure}[!ht]
\centering \psfull  \epsfig{file=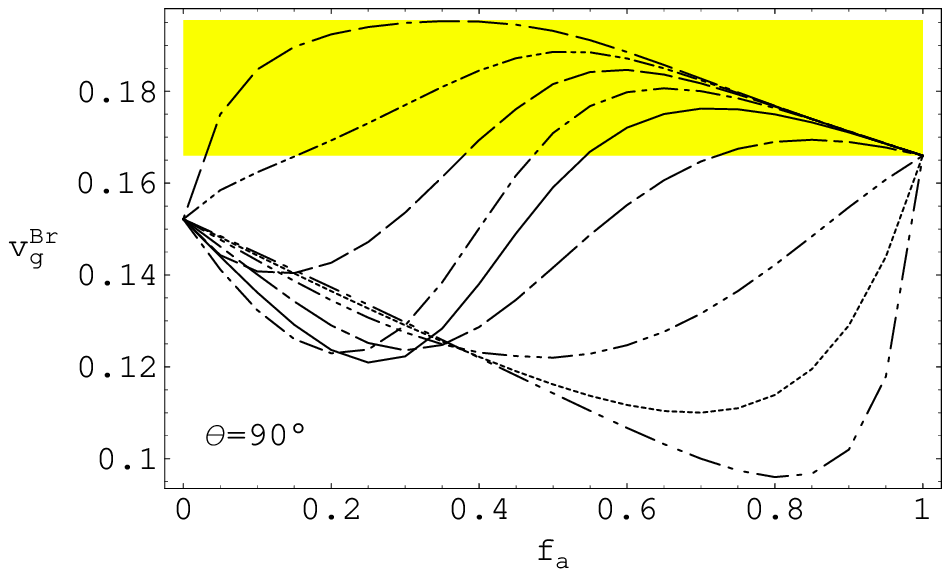,width=5.0in}
  \caption{\label{fig2d}
(d) Extraordinary wavevector angle $\theta = 90^\circ$.}
\end{figure}

\end{document}